\documentclass[useAMS,usenatbib]{mn2e}

\usepackage{graphicx}
\usepackage{amssymb}
\usepackage{journal_shortcuts}

\title{Metal transport by gas sloshing in M87}
\author[A. Simionescu et al.]{A. Simionescu$^{1,2}${\thanks{Einstein fellow, e-mail:
asimi@stanford.edu}}, N. Werner$^{1}${\thanks{Chandra/Einstein fellow}}, W. R. Forman$^3$, E. D. Miller$^4$, Y. Takei$^5$, \newauthor
H. B\"ohringer$^2$, E. Churazov$^{6,7}$, P. E. J. Nulsen$^3$\\
$^1$KIPAC, Stanford University, 452 Lomita Mall, Stanford CA 94305, USA\\
$^2$MPE, Gie{\ss}enbachstr. 1, 85748 Garching, Germany\\
$^3$Harvard-Smithsonian Center for Astrophysics, 60 Garden St., Cambridge MA 02138, USA\\
$^4$MIT, Kavli Institute for Astrophysics and Space Research, 77 Massachusetts Ave 37-582G, Cambridge MA 02139, USA\\
$^5$Institute of Space and Astronautical Science (ISAS), JAXA, 3-1-1 Yoshinodai, Sagamihara, Kanagawa 229-8510, Japan\\
$^6$MPA, Karl Schwarzschild Str. 1, 85748 Garching, Germany\\
$^7$Space Research Institute (IKI), Profsoyuznaya 84/32, Moscow 117810, Russia
}
\begin{document}

\date{Accepted . Received ; in original form 2009 September 21}

\pagerange{\pageref{firstpage}--\pageref{lastpage}} \pubyear{2009}

\maketitle

\label{firstpage}

\begin{abstract}

We present the results of an XMM-Newton mosaic covering the central $\sim200$~kpc of the nearby Virgo cluster. We focus on a strong surface brightness discontinuity in the outskirts of the brightest cluster galaxy, M87. Using both XMM-Newton and Suzaku, we derive accurate temperature and metallicity profiles across this feature and show that it is a cold front probably due to sloshing of the Virgo ICM. It is also associated with a discontinuity in the chemical composition. 
The gas in the inner, bright region of the front is $\sim$40\% more abundant in Fe than the gas outside the front, suggesting the important role of sloshing in transporting metals through the ICM.
For the first time, we provide a quantitative estimate of the mass of Fe transported by a cold front. This amounts to $\sim$6\% of the total Fe mass within the radial range affected by sloshing, significantly more than the amount of metals transported by the AGN in the same cluster core. 
The very low Fe abundance of only $\sim$0.2 solar immediately outside the cold front at a radius of 90 kpc suggests we are witnessing first-hand the transport of higher metallicity gas into a pristine region, whose abundance is typical of the cluster outskirts. 
The Mg/Fe and O/Fe abundance ratios remain approximately constant over the entire radial range between the centre of M87 and the faint side of the cold front, which requires the presence of a centrally peaked distribution not only for Fe but also for core-collapse type supernova products. 
This peak may stem from the star formation triggered as the BCG assembled during the protocluster phase.

\end{abstract}


\section{Introduction}
In the last decades, our understanding of the intricate phenomena in clusters of galaxies has advanced tremendously based on observations with the constantly improving generations of X-ray satellites. Clusters are now known to show complex morphologies and exciting physics on all scales and are thus among the most rewarding objects to target in the X-ray domain. Detailed spatial and spectral information about clusters can reveal important clues about the energetics of large-scale gas motions in the intracluster medium (ICM) and about the interaction between the ICM and the active galactic nucleus (AGN) in the central galaxy. Both AGN-ICM interactions in the form of shocks and bubbles \citep{Binney95,churazov2000,Boehringer02,Brueggen02,Birzan04,Forman05,Voit05,fabian2006,McNamara07} and gas "sloshing" in the cluster dark matter potential \citep{Markevitch01,Churazov03,markevitch2007} have been invoked, among other explanations, as possible ways to heat the gas in the centre of cool-core clusters, preventing the radiative cooling of the gas at high rates which should occur in the absence of heating but which is not observed \citep{Peterson01,Peterson03,Boehringer01}. Moreover, detailed spatial and spectral information about clusters is essential in constraining the chemical enrichment history of the ICM \citep{deplaa2007,Werner08rev,boehringer2009} and possible mechanisms responsible for the transport of heavy elements throughout the hot cluster medium \citep{Rebusco06,simionescu2008a,simionescu2008b}.

One of the best targets to perform in-depth studies of the ICM is M87 at the centre of the Virgo cluster. This is the nearest galaxy cluster ($\sim16$ Mpc), allowing us to resolve phenomena on small scales. M87 is among the brightest extragalactic X-ray sources in the sky, a guarantee for excellent spectral statistics and accurate temperature and metal abundance determinations within the technical capabilities of the detectors. M87 is among the best studied galaxies in the Universe and its central region has been the target of very deep observations with XMM-Newton, {\it Chandra}, and recently Suzaku. The hot gas atmosphere of M87 shows striking signs of AGN-ICM interaction, including an AGN-driven classical shock \citep{Forman05,Forman06,simionescu2007a} and clear enhancements in X-ray surface brightness associated with the radio lobes to the east and southwest of the core \citep{Feigelson87,bohringer1995,Belsole01,Young02,Forman05,Forman06}. Initial XMM-Newton observations showed that multi-temperature models including a cool gas component are needed to describe the spectra in these regions, also known as the E and SW X-ray arms \citep{Belsole01,Molendi02}. More recently, with deeper XMM-Newton data, a correlation was found between the percentage of cool gas and the metallicity in these regions, based on which \citet{simionescu2008a} concluded that the cool gas is metal-rich and that it was uplifted by the AGN \citep[following the scenario proposed by][]{Churazov01} from the central parts of the galaxy which had been enriched by stellar mass loss and supernova explosions. The energetics and metal-transport mechanisms associated with the AGN activity in the central parts of M87 have therefore been thoroughly investigated to date and have yielded important information about the physics of the AGN-ICM interaction. Here, we present the results of Suzaku and XMM-Newton observations extending the in-depth analysis to the outskirts of M87 to look for evidence for large-scale gas motions in the Virgo core and to bring gas sloshing and AGN-ICM interaction together in explaining the metal transport processes in the ICM.

We focus on a clearly observed surface brightness edge that lies 90 kpc (19$^\prime$) to the north of M87 and extends $140^{\circ}$ in azimuth in the archival ROSAT $\sim$30~ks image, its large scale making it an optimal target for Suzaku given the limitations related to its point-spread function. We show, using recent XMM-Newton and Suzaku data, that this edge is a ``cold front'', a contact discontinuity associated with gas sloshing in the ICM, typical of those found in many cool core clusters \citep{markevitch2007}. 
\citet{simionescu2007a} have previously detected a weaker counterpart cold-front $\sim 33$ kpc to the south-east of the core. The opposite and staggered placement of these two features is a typical example of sloshing as seen in simulations \citep{Tittley05,Ascasibar06}. As the gas sloshes back and forth in the gravitational potential, cooler and denser parcels of gas from central parts of the cluster can move and come into contact with the hotter outskirts creating ``cold-front'' signatures, where the temperature on the denser side of the surface brightness edge is lower, which is the opposite of what one expects from a shock front \citep{Markevitch01}. Since heavy elements in cooling flow clusters are concentrated toward the centre \citep[e.g.][]{degrandi2001,Leccardi08}, the displaced parcels of cool central gas are also more metal-abundant. Thus, the abundance should be discontinuous across these fronts, as long as sloshing occurs within the region with a strong gradient. In general, sloshing should spread the heavy elements from the centre outwards, providing, beside the central AGN, a mechanism for distributing and transporting the metals produced in the central galaxy into the ICM. 

In this paper, we discuss the abundance gradients for different heavy elements across the cold front to quantify the metal transport due to sloshing and to study the chemical enrichment histories of the gas on either side of the surface brightness edge.

\section{Data reduction}

\subsection{Suzaku}

On 2008 June 8, Suzaku observed two fields offset to the North by 19$^\prime$ and 33$^\prime$ with respect to the centre of M87. The effective exposure times were 59.7 ks for the pointing furthest away from the M87 centre (``M87Nfaint'') and 21.9 ks for the pointing with a smaller radial offset (``M87Nbright''). The distance between the Southern edge of the field-of-view of the M87Nfaint pointing and the surface brightness discontinuity was chosen to be $\sim 3^\prime$ (of the order of the Suzaku PSF) in order to avoid contamination from the bright side of the front into the faint-side spectra. 

\begin{figure}
\includegraphics[width=\columnwidth, bb=49 36 564 757]{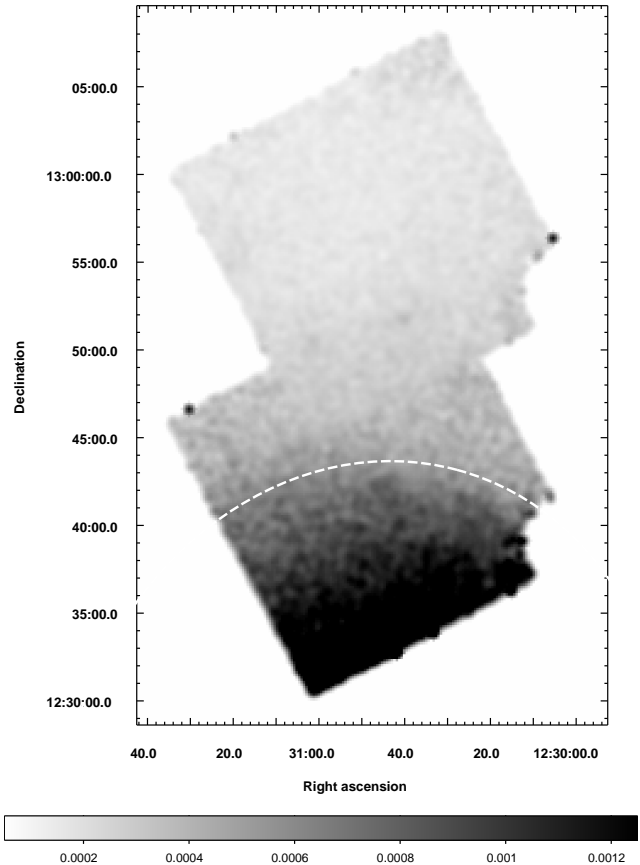}
\caption{Mosaic of the two Suzaku fields (exposure-corrected). The dashed line marks the approximate position of the outer cold front.}
\label{suzakuim}
\end{figure}

The data were reduced using the tools available in the HEAsoft package (version 6.7) to create a set of cleaned event lists with hot or flickering pixels removed and including only ASCA grades 0,2,3,4,6. The recommended filtering criteria were used to remove South Atlantic Anomaly (SAA) transits and times of low geomagnetic cut-off rigidity (COR), eliminating data with COR $\leq$ 6 GV. 
In addition, all data taken within 20$^\circ$ of the sun-lit Earth limb or 5$^\circ$ of the dark Earth limb were excised to reduce contamination from scattered solar X-ray flux. Data obtained with 3x3 and 5x5 editing modes were merged into a single event file. The lightcurves for both data sets after the initial cleaning do not show anomalous count rates either in the full or in the soft energy bands. Ray-tracing simulations based on the available ROSAT data show that the expected contamination by out-of-field-of-view emission is below 1.5\% in both pointings. An exposure-corrected flux image of the two combined Suzaku pointings is shown in Fig. \ref{suzakuim} (no background subtraction has been performed for the image processing). The cold front is clearly visible in the image.

For the spectral analysis, we subtracted the non X-ray background (NXB) obtained with \texttt{xisnxbgen} and then modeled the cosmic X-ray background (CXB) with 3 components: 
two thermal components to account for the local hot bubble (LHB) and for the Galactic halo (GH) emission, and a power-law to account for the integrated emission of unresolved point sources. The temperature of the LHB was frozen to 0.08~keV, that of the GH to 0.2~keV, and the power-law index to 1.41; unless otherwise noted, the spectrum normalisations were fixed based on the fluxes given by \citet{kuntz2000} for the thermal components and by \citet{deLuca04} for the power-law. We note that the region investigated is within approximately 0.2 of the virial radius of the Virgo cluster, thus the CXB level is still low compared to the emission from the ICM.

\subsection{XMM-Newton}

\begin{figure*}
\includegraphics[width=\textwidth,bb=36 237 577 555]{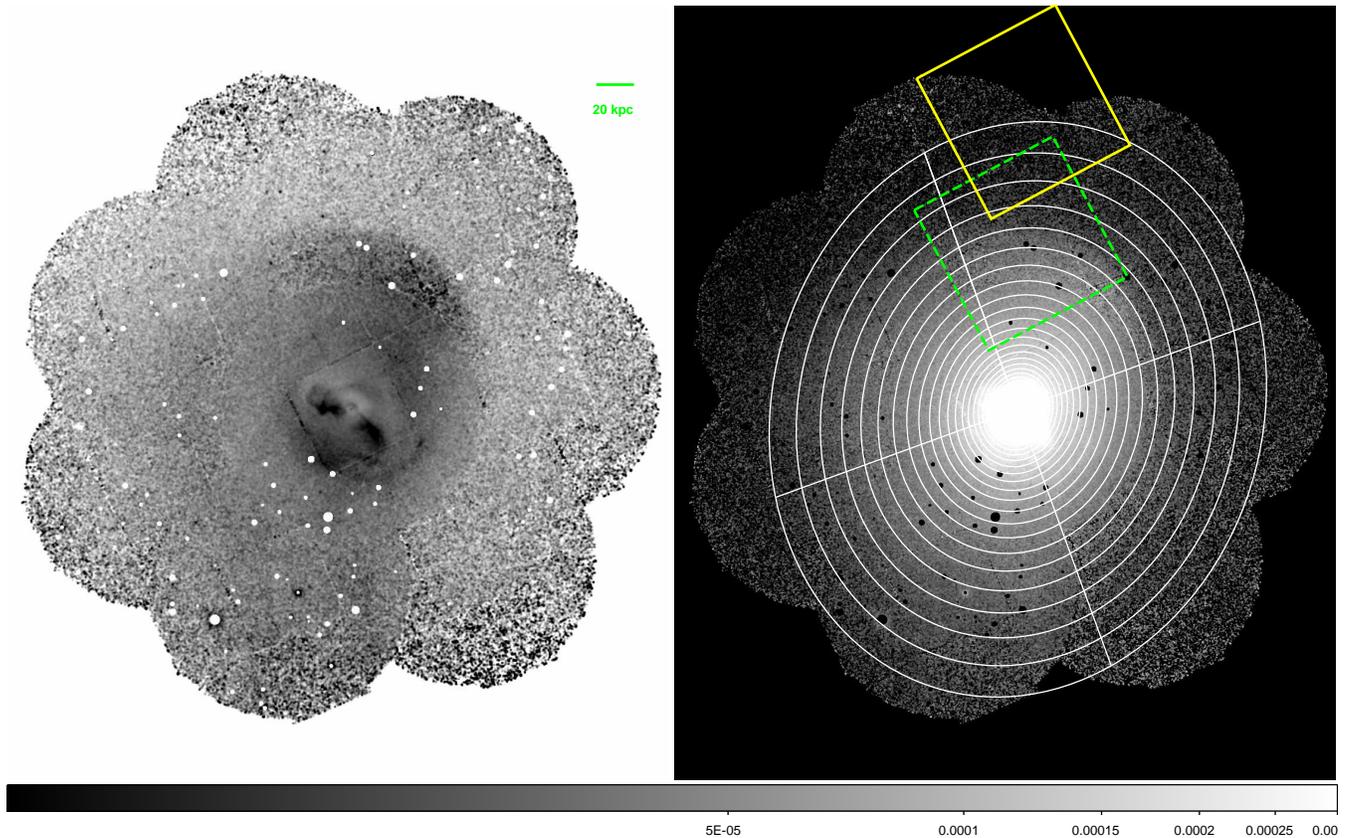}
\caption{XMM-MOS mosaic of M87. {\it Left:} Exposure corrected image divided by the best-fitting radially symmetric beta model. {\it Right:} Extraction regions used for the radial profiles in Sect. \ref{prof} and the two Suzaku fields overplotted on the exposure corrected image. Colorbar units are counts per second.}
\label{extrreg}
\end{figure*}

A series of 6 pointings with an average exposure time of 20~ks was performed with XMM-Newton in June 2008, with an additional seventh pointing of the same length completed on 2008 November 12. These seven pointings cover an annulus around the existing deep observations of the central part of M87 from 2000 June 19 and 2005 January 10, which had a combined effective exposure time of 120~ks. A combined exposure-corrected flux image of the MOS data from all the pointings is shown in Fig. \ref{extrreg}, together with the residuals obtained by dividing this image by a circularly symmetric beta model, with a best-fit core radius of $12.6^{\prime\prime}$ and $\beta=0.38$. Two surface brightness discontinuities, one at a radius of $\sim$33 kpc towards the SE and one at $\sim$90 kpc towards the NW are clearly seen. 

For data reduction we used the 8.0.0 version of the XMM-Newton Science Analysis System (SAS). Out-of-time events were subtracted from the EPIC/pn data using the standard SAS prescription for the extended full frame mode.
To remove flaring from soft protons, we used a two-stage cleaning of the event files. First, we extracted a light-curve in the hard energy band (10-12 keV for MOS1 and MOS2 and 12-14 keV for pn) using 100 second time bins and excluded the time periods in each observation when the count rate exceeded the mean by more than 3$\sigma$. After this first cleaning, we used the resulting event files to extract a light-curve in a broad energy band (0.3-10 keV) with 10-s time bins and again excluded the time intervals with count rates larger than the average by more than 3$\sigma$.
We then examined the soft band (0.3--0.5 keV) images of each MOS observation to look for chips with anomalously high flux in this energy range \citep{snowden2007}, and found that none of our observations were affected by this instrumental problem.

For the background subtraction, we used a combination of blank-sky maps from which point sources have been excised \citep{ReadPonman} and closed-filter observations. 
This is necessary because the instrumental background level of XMM-Newton is variable and increases with time. For each detector, we added to the corresponding blank sky background set a fraction of the closed filter data designed to compensate for the difference between the out of field of view (OoFoV) hard-band count rate in the observation and in the blank sky data \citep[for more details, see][]{simionescu2008b}.

\section{Temperature and iron abundance profiles}\label{prof}

We first present radial profiles across and away from the northwestern cold front in order to establish the projected temperature and Fe abundance jumps associated with this feature and compare the different trends towards the NW and SE from the core.

For the XMM-Newton data, the spectral extraction regions used were 90$^\circ$-wide sectors of elliptical annuli towards the NW and SE, respectively. The thickness of these annuli increases logarithmically with distance from the M87 centre, in order to balance the decreasing surface brightness. The exact regions used are shown in Fig. \ref{extrreg}. For the Suzaku data, we chose elliptical annuli with the same ellipticity and position angle as for the XMM-Newton data. For illustration, the ellipse delimiting the two Suzaku annuli immediately outside and immediately inside the outer cold front is over-plotted in Fig. \ref{suzakuim}. No sector selection was performed since the opening angle of the area covered by Suzaku is already limited (see green and yellow squares overplotted in Fig. \ref{extrreg}) and all annuli were fixed to have a width of 3$^\prime$, due to constraints from the Suzaku PSF. 

We use XSPEC 12.5.1 to model our spectra with a single-temperature plasma model in collisionally ionised equilibrium. Unless otherwise stated, we used the \texttt{apec} plasma model \citep{smith2001}. The determination of the Galactic absorption column density based on X-ray spectra can be easily biased by subtle underlying multi-temperature structure in the ICM \citep{simionescu2008b} or by uncertainties in the calibration of the oxygen edge around 0.5~keV and of Suzaku's contamination layer. We therefore opt to fix this parameter to $N_{\mathrm{H}}=2.0\times10^{20}$~cm$^{-2}$ \citep{Lieu96,kalberla2005}. 

Point sources identified using the X-ray images were excluded from the spectral analysis for the XMM-Newton spectra. No unusually bright point sources are present within the field of view of Suzaku. The spectra obtained by different detectors of the same satellite were fit simultaneously with their relative normalisations left as free parameters. This is mainly done in order to account for effective area variations due to XMM-Newton chip gaps, the failed CCD6 on MOS1, and the different locations of the field of view edges for MOS and pn. The instrument normalisations of the three XIS detectors agree to within 5\% for each region analyzed.
The MOS, pn, and XIS1 data were fit in the 0.4--7.0 keV band while for XIS0 and XIS3 we used the 0.4--10 keV band. The spectra were binned with a minimum of 30 counts per bin before fitting. 
The normalisation, temperature, and the abundances of O, Ne, Mg, Si, S, Ar, Ca, Fe, and Ni were left as free parameters. The abundances of the other elements heavier than He were fixed to 0.5 of the Solar value. For consistency with previous chemical enrichment studies of M87 with XMM-Newton by \citet{Matsushita03}, we adopt the solar abundances of \citet{Feldman92}. 

\begin{figure}
\includegraphics[width=\columnwidth,bb=40 144 380 690]{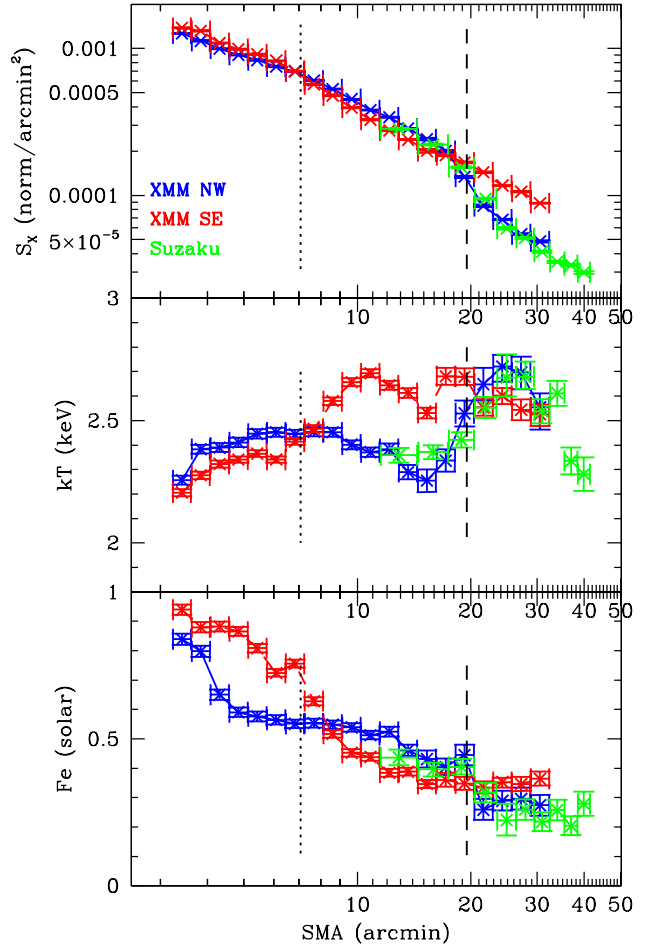}
\caption{X-ray surface brightness, temperature and Fe abundance profiles towards the NW and SE as a function of the average semi-major axis (SMA) of the elliptical annuli used as extraction regions. The positions of the two cold fronts at 33 and 90~kpc are marked with vertical dotted and dashed lines, respectively.}
\label{ktfeprof}
\end{figure}

Our spectral fitting results are shown in Fig. \ref{ktfeprof}. We find an excellent agreement between XMM-Newton and Suzaku in determining all the spectral parameters, despite the different background subtraction techniques employed. 
The X-ray surface brightness profile, obtained by dividing the spectrum normalisation from XSPEC by the area of the corresponding extraction region, shows a very clear discontinuity at $19^\prime$ ($\sim$90~kpc), the radius of the outer cold front. At the radius of the inner cold front, the surface brightness profile towards the SE crosses from being systematically higher to being systematically lower than the corresponding profile towards the NW. The jump however is less pronounced, and is only evident once the strong radial trend is removed using a symmetric model, as was done in Fig. \ref{extrreg}.
The temperature profile towards the SE rises beyond the inner cold front at $\sim$33~kpc discussed by \citet{simionescu2007a} and then becomes flat, while the projected temperature to the NW stays flat or decreases slightly in the radial range beyond 30~kpc and only shows a sharp rise at the cold front at $19^\prime$ ($\sim$90~kpc).
Therefore, from the temperature profiles, we can conclude that the surface brightness discontinuities both at 33 and 90~kpc are cold fronts rather than shocks, exhibiting a lower temperature of the gas on the denser side. After increasing outside of the 90~kpc cold front, the temperature seems to decrease again beyond 30$^\prime$. This is however most likely a feature of the general Virgo temperature profile rather than being associated with the event that caused the cold front. A more detailed deprojection analysis will be presented by Forman et al. (in prep). 

Typically two types of cold fronts have been discussed in the literature \citep[see][]{Owers09,markevitch2007} and two different mechanisms for producing them have been proposed. On one hand, cold fronts have been seen in merging clusters, where often the remnant core belonging to an infalling group or smaller cluster can still be identified. The best and most famous example here is the bullet cluster \citep{Markevitch02}. The cold front then exists at the interface between the cool, dense remnant core and the ICM of the cluster with which it is merging. This is clearly not the case for M87, which appears relatively relaxed apart from the X-ray arms and other small scale features in the very center (inner $\sim1^\prime$), all of which seem related to AGN-ICM interaction \citep[e.g.][]{Feigelson87,Matsushita02,Forman05,simionescu2008a}.
The most probable explanation for the opposite and radially staggered cold fronts in M87 thus is gas sloshing, similar to that proposed by \citet{Markevitch01} in A1795, where the gravitational potential of a cool core cluster can be perturbed by a past subcluster infall. Due to this disturbance, the central ICM is displaced and fronts are created where cooler and denser parcels of gas from the centre come into contact with the hotter outskirts. Numerical simulations show that the sloshing typically follows a spiral pattern, which can explain the placement of the two cold fronts in M87 at different radii on opposite sides from the centre \citep{Tittley05,Ascasibar06}. Using the mass profile of \citet{Matsushita02}, the Keplerian velocity is approximately 550 km/s for an orbit with a radius of 33~kpc and 850 km/s for a radius of 90~kpc. A rough estimate for the time needed for the two fronts to become out of phase by 180 degrees is thus $\pi/(\omega_{33}-\omega_{90})$ = 0.4 Gyr, where $\omega=v/r$ is the angular velocity at the corresponding radius. This estimate is likely to be low.

An interesting topic for further simulations in the case of M87 will be whether the large scale bulk motions due to ICM sloshing are strong enough to cause the observed bending of the AGN-inflated radio lobes and of the associated E and SW X-ray arms.

Apart from the marked increase in temperature beyond the surface brightness discontinuities at 33 and 90~kpc, another common feature is a sharp drop in Fe abundance outside each cold front. A discontinuity in the metal abundance profile is expected because the Fe abundance in Virgo, as in the vast majority of cooling core clusters, is centrally peaked \citep[Fig. \ref{ktfeprof}; see also e.g.][]{Leccardi08}. This means that the central gas displaced by sloshing is also richer in metals and thus sloshing contributes to transporting the heavy elements from the centre outwards.

To quantify this, we calculate the excess mass of Fe towards the SE behind the inner cold front (in an opening angle of $90^\circ$ between $\sim$14 and 33~kpc) and the excess mass of Fe towards the NW behind the outer cold front (in an opening angle of $90^\circ$ between $\sim$33 and 90 kpc). This is done simply by estimating the gas mass in each annulus based on the spectrum normalisation, multiplying by the Fe mass fraction determined from the Fe abundance, subtracting the corresponding value for the opposite sector, and finally summing over all annuli in the cited radial ranges. Note that this is likely an underestimate of the mass of Fe transported by the outer cold front. This is firstly because this front extends over an opening angle larger than $90^\circ$ (although it is less pronounced outside the region considered here). Secondly, the low Fe abundance outside the outer cold front suggests that the initial Fe profile to the NW before the metal transport was lower than estimated from taking the SE profile at corresponding radii. 

We obtain an excess Fe mass of $4.5\times10^6$ M$_\odot$ behind the inner cold front and $13.7\times10^6$ M$_\odot$ behind the outer cold front. 
This represents 31\% and 22\% of the average Fe mass per sector within the corresponding radial range for the inner (33~kpc) and outer (90~kpc) fronts, respectively. If the Fe mass averaged between the NW and SE sectors is representative also for each of the other two sectors towards the NE and SW, then the mass of Fe transported by sloshing represents $\sim$6--8\% of the total Fe mass within the considered radial ranges. 
Alternatively, if we relate the mass of Fe transported by the two cold fronts to the total Fe masses between 14--33 and 33--90~kpc obtained using the cumulative Fe mass profile of M87 from \citep{bohringer2004}, we obtain very similar ratios of $\sim$6--9\%.

It is noteworthy that both the large-scale 90~kpc cold front and the inner cold front seem to transport a very similar fraction of the mass of Fe available within the radial region affected by the sloshing.
By comparison, the Fe mass uplifted by the AGN in the bright X-ray arms is only $1.5\times10^6$ M$_\odot$ \citep{simionescu2008a}. Note that this is only an estimate of the current amount of metals being displaced by either process - depending on the strength of the mixing within the ICM, a fraction of these metals may eventually fall back to their original position within the cluster.

While it was expected that the sloshing that causes cold fronts is an important means of transporting metals in the ICM and that this would cause the Fe abundance to be discontinuous at the surface brightness jump towards the NW, the strength of this feature is surprising. Immediately outside the cold front at 90 kpc, we reach a very low, flat abundance plateau. The average abundance here is $0.25\pm0.02$ solar (see also next section) relative to the solar units of \citet{Feldman92}, which translates to only $0.17\pm0.01$ solar in the units of \citet{Angr}. This is as low as the expected metallicity close to the virial radius in other clusters \citep[0.2 solar,][]{fujita2007}. Such low Fe abundances are rarely measured in the ICM. For a sample of nearby galaxy clusters observed with XMM-Newton, \citet{Leccardi08} find that the average metallicity profile keeps decreasing out to very large radii, reaching $\sim$0.2 solar only at around 0.3--0.4 r$_{180}$. A similar trend is shown by the Fe abundance profiles measured with Suzaku \citep[e.g.][]{Sato09}. 
For an average temperature of 2.5~keV, the virial radius of the Virgo cluster based on the scaling relations of e.g. \cite{Arnaud05} is $r_{200}\sim$ 1.2~Mpc, which makes it very surprising to see gas with such a low metallicity outside a radius of only 90~kpc. 
Towards the SE, the metallicities at corresponding radii beyond the outer cold front are systematically higher, closer to what one would expect at these relatively small distances from the cluster centre. The likely scenario is that sloshing in the more distant past has influenced the abundances of the gas towards the SE and, because Virgo is a relatively young and unrelaxed object, we are only now witnessing the dispersion of higher metallicity gas into the pristine region towards the NW, whose abundance is that typical of the cluster outskirts. 

\section{Metal abundance ratios and supernova enrichment history}

It is believed that the peak in Fe abundance associated with cool core clusters is produced by type Ia supernova ejecta in the brightest cluster galaxy (BCG) which is spatially coincident with this peak \citep{degrandi2004,bohringer2004}. While SN~Ia have a longer delay time and still occur in these BCGs, which have typically old stellar populations, core-collapse supernovae (SN$_{\rm CC}$) usually explode only within a short time after a star formation event and would thus not be expected to enhance abundances significantly in cool cores. Elements which are predominantly produced by SN$_{\rm CC}$, such as O, Ne and Mg, or even intermediate elements produced by both SN$_{\rm CC}$ and SN~Ia such as Si and S, are therefore expected to be less abundant compared to Fe in the vicinity of the BCG. Whether or not such trends in the e.g. O/Fe or Si/Fe ratios are in fact observed is, however, a matter of debate in recent literature. \citet{finoguenov2000} show an increase of Si/Fe with radius in a sample of clusters observed with ASCA. While the presence of such a gradient was confirmed, more recently, for the case of galaxy groups \citep{Rasmussen09}, the {\it constant} Si/Fe and increasing O/Fe between the centre and outskirts of M87 prompted \citet{finoguenov2002} to propose a diversity of SN~Ia: faint SN~Ia with less complete burning and higher Si/Fe yields would contribute to the enrichment of the central parts of the ICM, while brighter SN~Ia with lower Si/Fe yields would dominate the enrichment in the outskirts. A strong increase in O/Fe with radius claimed by \citet{tamura2004} has not been confirmed by recent analyses of deeper cluster observations, which show no significant increase in O/Fe within 0.1 r$_{200}$ \citep{simionescu2008b}. The latest Suzaku results also indicate constant Mg/Fe, Si/Fe, and S/Fe abundance ratios with radius and only a very weak, low-significance increase in O/Fe out to as far as 0.4 r$_{180}$ for several objects which have been studied in detail \citep{sato2007b,Sato09}.   

Trends in abundance ratios across cold fronts are an important missing piece of the puzzle in this ongoing debate. This is because sloshing effectively brings into contact gas which has been near the BCG and ``pristine'' gas at large radii, which has not been influenced (or has been influenced much less) by the BCG at late times. If the BCG has a large impact on the metal budget of the gas in its vicinity, this should be reflected in marked differences between the chemical enrichment patterns of the gas on either side of the cold front.
To test this, we focus here on the northwestern cold front at 90~kpc, for which the good energy resolution of the Suzaku observations allows more precise determinations of the metal abundances than possible with other data sets \citep[see previous work by e. g. ][for the case of A496 with {\it Chandra}]{Dupke03,Dupke07}. 

Since precise determinations of the metal abundances other than Fe require significantly more counts, we must substantially increase the sizes of the extraction regions. We thus can only determine one value of the O and Mg abundances for each side of the cold front, rather than being able to produce a radial profile. For this, we fitted in parallel the Suzaku spectra extracted from the first 3 and last 6 ellipses used to create the radial profiles in Fig. \ref{ktfeprof} to determine the metal abundances on the bright and faint side of the cold front, respectively. The temperature and normalisation of each ellipse spectrum was free in the fit, while all elemental abundances were coupled between the data sets in each case. This ensures that we account for variations in temperature across the large regions sampled by the spectra, which would otherwise bias the abundance measurements. Note that allowing the Fe abundance also to vary between the different elliptical extraction regions does not change the measured abundances for the other elements. The results are shown in Table \ref{abund}.

\begin{table}
\caption{Metal abundances on the bright and faint sides of the cold front.
The Suzaku spectra extracted from the first 3 and last 6 ellipses used to create the radial profiles were fit in parallel with the temperature and normalisations left free (3T/6T model) in order to determine the metal abundances on the bright and faint side of the cold front, respectively. The errors are given at the $\Delta \chi^2$=1 confidence level.}
\begin{center}
\begin{tabular}{l|cc}
\hline
\hline
Parameter 	& bright side 3T     & faint side 6T  \\
\hline
O	        & $0.30\pm0.07$      & $0.17\pm0.07$ \\
Ne              & $0.42\pm0.07$      & $0.31\pm0.08$ \\
Mg            	& $0.34\pm0.07$      & $0.22\pm0.07$ \\  
Si              & $0.36\pm0.04$      & $0.16\pm0.04$ \\
S               & $0.39\pm0.05$      & $0.23\pm0.05$\\
Ar              & $0.55\pm0.10$      & $0.45\pm0.12$\\
Ca            	& $0.63\pm0.16$      & $0.60\pm0.18$ \\     
Fe              & $0.41\pm0.02$      & $0.25\pm0.02$\\
Ni            	& $0.35\pm0.14$      & $0.28\pm0.16$  \\      
\hline
$\chi^2$/d.o.f. & 2730/2696 & 3848/3678 \\
\hline 
\end{tabular}
\end{center}
\label{abund}
\end{table}

\begin{figure}
\includegraphics[width=\columnwidth, bb= 18 144 592 718]{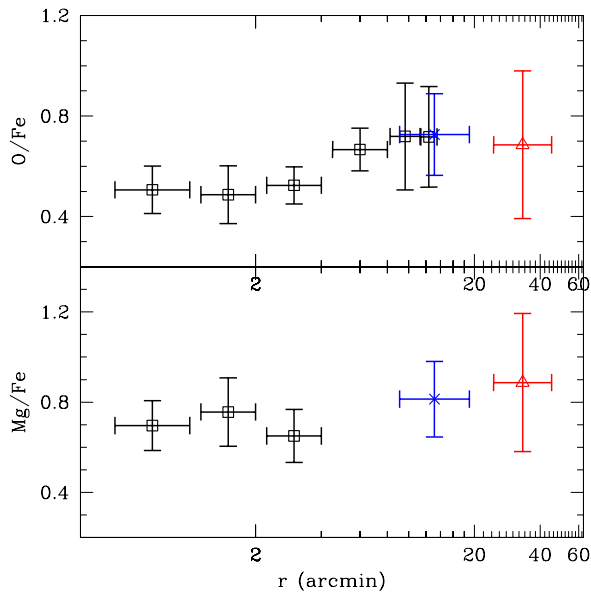}
\caption{O/Fe and Mg/Fe abundance profiles. In black, values determined for several central annuli by \citet{Matsushita03} with XMM-Newton. In blue, the results for the bright side and in red, for the faint side of the cold front.}
\label{omgfeprof}
\end{figure}

We caution that the abundance of any element produced predominantly by SN$_{\rm CC}$ is unfortunately very difficult to determine from the X-ray spectra. The Ne X line is blended with the strong Fe-L complex between 0.9--1.2 keV, which is very sensitive to the exact modeling of the temperature structure of the emitting gas. The O determination is sensitive to the normalisation of the 0.2~keV galactic halo component, which is difficult to determine exactly because the soft X-ray background around M87 is patchy. Raising or lowering this normalisation by a factor of 2 changes the best-fitting O abundance on the faint side of the cold front by as much as 40\%, while Mg changes only by 14\% and the temperature by less than 0.02~keV. The Mg line is also blended with Fe-L emission and, although the Fe-L line emission around the Mg line is much weaker than that surrounding the Ne line, different spectral codes disagree on its modeling. The current collisional ionization equilibrium (CIE) plasma model implemented in SPEX \citep{kaastra1996} gives Mg abundances lower by a factor of 2 compared to the \texttt{vapec} model, which provides a better fit to the Mg/Fe-L line blend \citep[e.g.][]{Matsushita03}, motivating our choice of the latter plasma model for the present work. An improved version of SPEX containing an updated Fe-L line library, which is yet to be released, provides better agreement with \texttt{vapec} (J. de Plaa, private communication; see Table \ref{mekalapec}). The temperature structure moreover plays an important role in determining the elemental abundances, including Mg. 
If the metal abundances are obtained by fitting the integrated spectra of the regions in front of and beyond the cold front with single temperature models, there is an indication of a relative increase in Mg/Fe on the faint side compared to the bright side of the surface brightness jump, despite the systematic offsets in determining the absolute Mg/Fe between the different plasma models. However, accounting for the temperature structure by allowing the temperature to vary across several rings with fixed abundances shows that this trend is only an artefact.

\begin{table}
\caption{Systematics in determining the Mg/Fe ratio.
The 3T/6T model refers to fitting the first 3 and last 6 ellipses used to create the radial profiles in parallel with the temperature and normalisations left free.
}
\begin{center}
\begin{tabular}{l|cc}
\hline
\hline
Mg/Fe	       & bright side	  & faint side  \\
\hline
vapec 3T/6T	  & $0.81\pm0.17$      & $0.88\pm0.29$ \\
vapec 1T	  & $0.77\pm0.15$      & $1.14\pm0.32$ \\  
spex 1T 	  & $0.42\pm0.15$      & $0.68\pm0.32$ \\
new spex 1T	  & $0.97\pm0.18$      & $1.28\pm0.40$\\
\hline 
\end{tabular}
\end{center}
\label{mekalapec}
\end{table}

The results presented in Table \ref{abund} show a decrease both in the Fe abundance and in the abundances of all elements predominantly produced by SN$_{\mathrm{CC}}$ across the cold front. Although the individual significances of the changes in the O, Ne and Mg abundances are small due to the large statistical uncertainties, the fact that all three abundances decrease by a similar relative amount (26--43\%), which is approximately the same as the relative change in the Fe abundance (40\%), is a convincing indication that the measured differences are reliable. Adding in quadrature the significances of the variations in O, Ne and Mg across the cold front suggests a change in the abundance of SN$_{\mathrm{CC}}$ products at the 90\% confidence level. This change requires the presence of a centrally peaked distribution not only for Fe but also for O, Ne, and Mg, and rules out models in which SN$_{\mathrm{CC}}$ products are uniformly distributed throughout the ICM, while only SN~Ia products show a radial gradient. 

Because all elemental abundances decrease by very similar relative amounts, we find no significant change in either Mg/Fe, Ne/Fe or O/Fe across the cold front. 
Comparing the Mg/Fe and O/Fe ratios on either side of the cold front with those measured at smaller radii by \citet{Matsushita03} using XMM-Newton data (also using the \texttt{vapec} model), we find a very slow increase across the entire radial range.
The data are consistent with no increase in Mg/Fe from the very centre of the galaxy out to roughly 210 kpc (45$^\prime$), as shown in Fig. \ref{omgfeprof}. 
\citet{Doherty09} show that the stellar halo of M87 truncates at an average radius of $\sim$150 kpc, therefore we are clearly probing the chemical enrichment history in regions both inside and beyond the sphere of influence of the galaxy. The very low Fe abundance outside the cold front (beyond $20^\prime$ in Fig. \ref{ktfeprof}) also strongly suggests that we are probing the chemical enrichment of the abundance floor at which the ICM metallicity is expected to flatten out at large radii.
The roughly constant Mg/Fe and O/Fe abundance ratios indicate similar chemical compositions and, consequently, enrichment histories in all these regions, although within the large statistical errors we cannot rule out a recent influence of SN~Ia in M87 on the metallicity budget of the Virgo ICM. 

Assuming the core-collapse model yields of \citet{nomoto2006} with solar metallicity of the progenitor,
the best-fitting Mg/Fe ratio in the centre of M87 (based on the average of the three data points from \citealt{Matsushita03}) corresponds to 27\% of all supernova explosions contributing to the chemical enrichment being SN~Ia, while the best-fitting Mg/Fe ratio on the faint side of the cold front suggests a relative contribution of $N_{\rm SN~Ia}/(N_{\rm SN~Ia}+N_{\rm SNcc}) \sim$20\% (independent of the assumed SN~Ia yield model). An enrichment of the gas on the faint side of the cold front (beyond 90~kpc) purely by SN$_{\mathrm{CC}}$ without any contributions by SN~Ia would imply a Mg/Fe ratio of 2.6 solar and is ruled out at the 5.8$\sigma$ level. 

The total mass of Fe contained within a 90~kpc radius from the centre of M87 is approximately $2.3\times10^8$ M$_\odot$ \citep{bohringer2004}. If 27\% of all supernovae in the core region are SN~Ia (based on the Mg/Fe ratio discussed above) and each SN~Ia produces $\sim$0.8 M$_\odot$ of Fe \citep{iwamoto1999}, while the rest are core collapse supernovae \citep[which, for a solar metallicity of the progenitor and assuming a Salpeter IMF, produce approximately 0.09 M$_\odot$ of Fe per explosion,][]{nomoto2006}, then $2.2\times10^8$ SN~Ia and $6\times10^8$ SN$_{\mathrm{CC}}$ are needed to produce all the metals observed in this region. 
For a Salpeter initial mass function (IMF) between 0.1 and 100 M$_\odot$, approximately one star with $M>8$ M$_\odot$ which will explode as a SN$_{\mathrm{CC}}$ is expected per 100 solar masses of stars formed. For a B-band mass to light ratio of around 8 \citep{Kronawitter00} and a blue-band luminosity of $10^{11}L_{B\odot}$ corresponding to the absolute B magnitude of M87 of -22.14 cited by \citet{Peletier90}, the stellar mass of M87 is $8\times10^{11}$ M$_\odot$ and $8\times10^9$ SN$_{\mathrm{CC}}$ would have exploded as the galaxy was forming, 13 times more than the $6\times10^8$ SN$_{\mathrm{CC}}$ needed to produce the metals in the central 90~kpc from the M87 centre. The central Mg peak can thus be explained if only less than 10\% of the SN$_{\mathrm{CC}}$ products from the galaxy formation phase are retained rather than being uniformly distributed into a flat abundance profile.
The expected present SN~Ia rates in elliptical galaxies are still relatively uncertain, from $0.18\pm0.06$ SNU \citep{Cappellaro99} to $0.36^{+0.22}_{-0.14}\pm0.02$ SNU \citep{Sharon07} or $0.28^{+0.11}_{-0.08}$ SNU \citep{Mannucci08}, where 1 SNU is equal to one supernova explosion per century per $10^{10}L_{B\odot}$ of the galaxy. 
Since redshift z=1, corresponding to a look-back time of 7.7 Gyr, between $10^8$ and $4.6\times10^8$ SN~Ia could have enriched the centre of M87 within the current uncertainties, between half to twice the amount needed in total. It is therefore possible that a fraction of the SN~Ia products in the central peak stem from the proto-cluster phase rather than recent SN~Ia in M87, although that requires the SN~Ia rates to be on the lower end of the measured confidence intervals and not to have been larger in the past, as predicted by e.g. \citet{Renzini93}; using 0.28 SNU, the most recent measurement of \citet{Mannucci08}, exactly the required number of SN~Ia, $2.2\times10^8$, would explode since z=1, thus all the Fe would be produced only by the central galaxy at late times. Another possibility to explain the constant Mg/Fe ratio in this case is that some of the Fe outside the metal peak was produced recently by SN~Ia explosions of intra-cluster stars. For a diffuse intra-cluster light with an average apparent magnitude of 27 mag/arcsec$^2$ \citep{Mihos05,Kormendy09}, 30\% of the Fe outside the 90~kpc cold front and within the radius covered by the XMM-Newton mosaic could have been produced by SN~Ia since z=1, for a SN~Ia rate of 0.28 SNU.

\begin{figure}
\includegraphics[width=\columnwidth,bb=18 144 592 718]{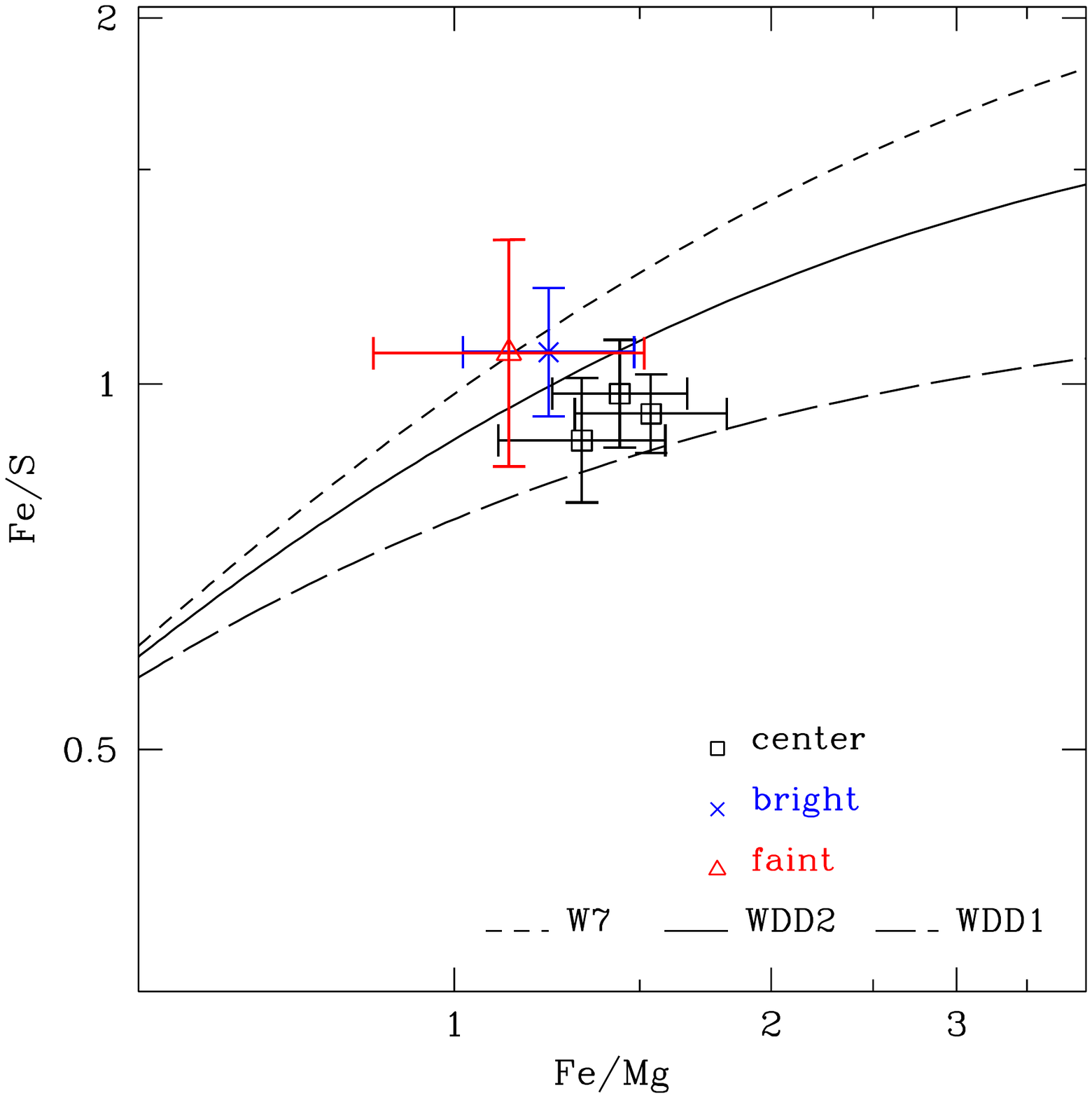}
\caption{Fe/S vs. Fe/Mg abundance ratios on the two different sides of the cold front (this work) and in the centre of M87 \citep{Matsushita03}. Overplotted are supernova enrichment models obtained by combining various relative numbers of SN~Ia \citep[using the WDD1, WDD2 and W7 yield models of ][]{iwamoto1999} and SN$_{\mathrm{CC}}$ \citep[whose yields are modeled based on][assuming a solar metallicity of the progenitor and averaged over the Salpeter IMF]{nomoto2006}.}
\label{femgs}
\end{figure}

We plot in Fig. \ref{femgs} the measured Fe/S vs. Fe/Mg ratios for the M87 centre \citep{Matsushita03} and on either side of the observed cold front using our measured values. The Fe/Mg, as discussed above, is an important diagnostic of the relative contribution of SN~Ia to SN$_{\mathrm{CC}}$. Using this ratio and assuming the same core-collapse model yields as above, we can determine the amount of Si-group elements produced by the SN$_{\mathrm{CC}}$ and consequently the ratio of Si-group elements to Fe produced by SN~Ia, which is an important diagnostic for the explosion mechanism of the latter. We overplot in Fig. \ref{femgs} three different  abundance ratio models based on the deflagration (W7) and delayed detonation (WDD1, WDD2) SN~Ia yields of \citet{iwamoto1999}. We used S as a representative element for the Si-group because the Si abundance is more easily biased by calibration issues around the Si edge and by modeling the underlying Si instrumental line. We find marginal evidence supporting the scenario proposed by \citet{finoguenov2002}, in that the data points at larger radii favor a W7 model whereas the abundance ratios at the centre of M87 favor the WDD1/WDD2 SN~Ia yields.

\section{Conclusions}

Two cold fronts are observed in M87: one towards the SE at a radius of $\sim$33~kpc, and one towards the NW with a radius of $\sim90$~kpc. The opposite and staggered placement of these cold fronts, as well as the absence of a remnant subcluster, make sloshing the most viable explanation for the presence of these cold fronts. The ICM distribution in and around M87 is very similar to results obtained from numerical simulations of sloshing in cool core clusters \citep{Tittley05,Ascasibar06}. For the mass profile of M87, the time needed for two fronts at these radii to become out of phase by 180 degrees is approximately 0.4 Gyr.

We present accurate projected temperature profiles which show that the temperature is higher on the X-ray fainter side of each cold front, opposite from what would be expected in the case of shock fronts. We also show that the metallicity drops sharply outside each cold front.
Such a discontinuity in the metal abundance profile is expected because sloshing displaces the central, cooler, more metal-rich gas and brings it into contact with hotter, more metal-poor ICM in the outskirts.
In this way, sloshing also contributes to transporting the heavy elements from the centre outwards.

We estimate that the inner cold front transports $4.5\times10^6$ M$_\odot$ of Fe, while for the outer cold front the total mass of Fe transported amounts to $13.7\times10^6$ M$_\odot$. These masses represent in each case the same fraction ($\sim$6--8\%) of the total Fe mass available within the radial range affected by sloshing, which is much larger than the amount of Fe transported by AGN-related processes, amounting to only $1.5\times10^6$ M$_\odot$ of Fe.

We find no strong indications for a change in the Mg/Fe and O/Fe ratios across a large radial range including the outer cold front, given a proper modeling of the temperature distribution within the extraction regions. 
This requires the presence of a centrally peaked distribution not only for Fe but also for O, Ne, and Mg, and rules out models in which SN$_{\mathrm{CC}}$ products are uniformly distributed throughout the ICM, while only SN~Ia products show a radial gradient. 
Since stellar mass loss in the central galaxy typically contributes only small amounts of heavy elements to the ICM metal budget,
the peaks in the distribution of SN$_{\mathrm{CC}}$ products are likely to stem from star formation, triggered as the BCG assembled during the protocluster phase, as proposed by \citet{simionescu2008b}. The central Mg peak can be explained if only less than 10\% of the SN$_{\mathrm{CC}}$ products from the galaxy formation phase are retained rather than being uniformly distributed into a flat abundance profile.

The very low Fe abundance of only $0.25\pm0.02$ solar \citep[$0.17\pm0.01$ relative to][]{Angr} suggests that the gas outside the cold front is representative of the pristine outskirts of typical clusters. This means that we are currently observing the very mechanism which spreads the metals into these pristine regions, broadening the metal distribution to that usually seen in relaxed clusters. Moreover, we are able to determine the chemical composition of this pristine gas, and find that probably around 20\% of all supernovae contributing to its enrichment are SN~Ia, that the metal abundance ratios of this gas are very similar to those near the centre of the BCG, and that an enrichment by SN$_{\mathrm{CC}}$ products only can be excluded with a very high significance.

\section*{Acknowledgments}
The authors would like to thank J. de Plaa, A. Finoguenov, J. S. Kaastra, E. Roediger, M. Br\"uggen, C. Jones, A. Baldi and H. Matsumoto for helpful discussion. 
Support for this work was provided by NASA through Einstein Postdoctoral Fellowship grants number PF9-00070 and PF8-90056 awarded by the Chandra X-ray Center, which is operated by the Smithsonian Astrophysical Observatory for NASA under contract NAS8-03060. We acknowledge NASA grant NNX08AZ88G.
AS is grateful for the hospitality of the Harvard-Smithsonian CfA. This work is based on observations obtained with XMM-Newton, an ESA science mission with instruments and contributions directly funded by ESA member states and the USA (NASA). The authors thank the Suzaku operation team and Guest Observer Facility, supported by JAXA and NASA. HB acknowledges support by the DfG Schwerpunkt Programm SPP 1177 and the Transregio Program TR33, the Dark Universe.

\bibliographystyle{mn2e}
\bibliography{bibliography,clustersnewest,sloshing}

\begin{thebibliography}{}

\bibitem[\protect\citeauthoryear{{Anders} \& {Grevesse}}{{Anders} \&
  {Grevesse}}{1989}]{Angr}
{Anders} E.,  {Grevesse} N.,  1989, \gca, 53, 197

\bibitem[\protect\citeauthoryear{{Arnaud}, {Pointecouteau} \& {Pratt}}{{Arnaud}
  et~al.}{2005}]{Arnaud05}
{Arnaud} M.,  {Pointecouteau} E.,    {Pratt} G.~W.,  2005, \aap, 441, 893

\bibitem[\protect\citeauthoryear{{Ascasibar} \& {Markevitch}}{{Ascasibar} \&
  {Markevitch}}{2006}]{Ascasibar06}
{Ascasibar} Y.,  {Markevitch} M.,  2006, \apj, 650, 102

\bibitem[\protect\citeauthoryear{{Belsole}, {Sauvageot}, {B{\"o}hringer},
  {Worrall}, {Matsushita}, {Mushotzky}, {Sakelliou}, {Molendi}, {Ehle},
  {Kennea}, {Stewart} \& {Vestrand}}{{Belsole} et~al.}{2001}]{Belsole01}
{Belsole} E.,  {Sauvageot} J.~L.,  {B{\"o}hringer} H.,  {Worrall} D.~M.,
  {Matsushita} K.,  {Mushotzky} R.~F.,  {Sakelliou} I.,  {Molendi} S.,  {Ehle}
  M.,  {Kennea} J.,  {Stewart} G.,    {Vestrand} W.~T.,  2001, \aap, 365, L188

\bibitem[\protect\citeauthoryear{{Binney} \& {Tabor}}{{Binney} \&
  {Tabor}}{1995}]{Binney95}
{Binney} J.,  {Tabor} G.,  1995, \mnras, 276, 663

\bibitem[\protect\citeauthoryear{{B{\^i}rzan}, {Rafferty}, {McNamara}, {Wise}
  \& {Nulsen}}{{B{\^i}rzan} et~al.}{2004}]{Birzan04}
{B{\^i}rzan} L.,  {Rafferty} D.~A.,  {McNamara} B.~R.,  {Wise} M.~W.,
  {Nulsen} P.~E.~J.,  2004, \apj, 607, 800

\bibitem[\protect\citeauthoryear{{B{\"o}hringer}, {Belsole}, {Kennea},
  {Matsushita}, {Molendi}, {Worrall}, {Mushotzky}, {Ehle}, {Guainazzi},
  {Sakelliou}, {Stewart}, {Vestrand} \& {Dos Santos}}{{B{\"o}hringer}
  et~al.}{2001}]{Boehringer01}
{B{\"o}hringer} H.,  {Belsole} E.,  {Kennea} J.,  {Matsushita} K.,  {Molendi}
  S.,  {Worrall} D.~M.,  {Mushotzky} R.~F.,  {Ehle} M.,  {Guainazzi} M.,
  {Sakelliou} I.,  {Stewart} G.,  {Vestrand} W.~T.,    {Dos Santos} S.,  2001,
  \aap, 365, L181

\bibitem[\protect\citeauthoryear{{B{\"o}hringer}, {Matsushita}, {Churazov},
  {Finoguenov} \& {Ikebe}}{{B{\"o}hringer} et~al.}{2004}]{bohringer2004}
{B{\"o}hringer} H.,  {Matsushita} K.,  {Churazov} E.,  {Finoguenov} A.,
  {Ikebe} Y.,  2004, \aap, 416, L21

\bibitem[\protect\citeauthoryear{{B{\"o}hringer}, {Matsushita}, {Churazov},
  {Ikebe} \& {Chen}}{{B{\"o}hringer} et~al.}{2002}]{Boehringer02}
{B{\"o}hringer} H.,  {Matsushita} K.,  {Churazov} E.,  {Ikebe} Y.,    {Chen}
  Y.,  2002, \aap, 382, 804

\bibitem[\protect\citeauthoryear{{B\"ohringer}, {Nulsen}, {Braun} \&
  {Fabian}}{{B\"ohringer} et~al.}{1995}]{bohringer1995}
{B\"ohringer} H.,  {Nulsen} P.~E.~J.,  {Braun} R.,    {Fabian} A.~C.,  1995,
  \mnras, 274, L67

\bibitem[\protect\citeauthoryear{{B{\"o}hringer} \& {Werner}}{{B{\"o}hringer}
  \& {Werner}}{2009}]{boehringer2009}
{B{\"o}hringer} H.,  {Werner} N.,  2009, \aapr, pp 11--+

\bibitem[\protect\citeauthoryear{{Br{\"u}ggen} \& {Kaiser}}{{Br{\"u}ggen} \&
  {Kaiser}}{2002}]{Brueggen02}
{Br{\"u}ggen} M.,  {Kaiser} C.~R.,  2002, \nat, 418, 301

\bibitem[\protect\citeauthoryear{{Cappellaro}, {Evans} \&
  {Turatto}}{{Cappellaro} et~al.}{1999}]{Cappellaro99}
{Cappellaro} E.,  {Evans} R.,    {Turatto} M.,  1999, \aap, 351, 459

\bibitem[\protect\citeauthoryear{{Churazov}, {Br{\"u}ggen}, {Kaiser},
  {B{\"o}hringer} \& {Forman}}{{Churazov} et~al.}{2001}]{Churazov01}
{Churazov} E.,  {Br{\"u}ggen} M.,  {Kaiser} C.~R.,  {B{\"o}hringer} H.,
  {Forman} W.,  2001, \apj, 554, 261

\bibitem[\protect\citeauthoryear{{Churazov}, {Forman}, {Jones} \&
  {B{\"o}hringer}}{{Churazov} et~al.}{2000}]{churazov2000}
{Churazov} E.,  {Forman} W.,  {Jones} C.,    {B{\"o}hringer} H.,  2000, \aap,
  356, 788

\bibitem[\protect\citeauthoryear{{Churazov}, {Forman}, {Jones} \&
  {B{\"o}hringer}}{{Churazov} et~al.}{2003}]{Churazov03}
{Churazov} E.,  {Forman} W.,  {Jones} C.,    {B{\"o}hringer} H.,  2003, \apj,
  590, 225

\bibitem[\protect\citeauthoryear{{De Grandi}, {Ettori}, {Longhetti} \&
  {Molendi}}{{De Grandi} et~al.}{2004}]{degrandi2004}
{De Grandi} S.,  {Ettori} S.,  {Longhetti} M.,    {Molendi} S.,  2004, \aap,
  419, 7

\bibitem[\protect\citeauthoryear{{De Grandi} \& {Molendi}}{{De Grandi} \&
  {Molendi}}{2001}]{degrandi2001}
{De Grandi} S.,  {Molendi} S.,  2001, \apj, 551, 153

\bibitem[\protect\citeauthoryear{{De Luca} \& {Molendi}}{{De Luca} \&
  {Molendi}}{2004}]{deLuca04}
{De Luca} A.,  {Molendi} S.,  2004, \aap, 419, 837

\bibitem[\protect\citeauthoryear{{de Plaa}, {Werner}, {Bleeker}, {Vink},
  {Kaastra} \& {M{\'e}ndez}}{{de Plaa} et~al.}{2007}]{deplaa2007}
{de Plaa} J.,  {Werner} N.,  {Bleeker} J.~A.~M.,  {Vink} J.,  {Kaastra} J.~S.,
    {M{\'e}ndez} M.,  2007, \aap, 465, 345

\bibitem[\protect\citeauthoryear{{Doherty}, {Arnaboldi}, {Das}, {Gerhard},
  {Aguerri}, {Ciardullo}, {Feldmeier}, {Freeman}, {Jacoby} \&
  {Murante}}{{Doherty} et~al.}{2009}]{Doherty09}
{Doherty} M.,  {Arnaboldi} M.,  {Das} P.,  {Gerhard} O.,  {Aguerri} J.~A.~L.,
  {Ciardullo} R.,  {Feldmeier} J.~J.,  {Freeman} K.~C.,  {Jacoby} G.~H.,
  {Murante} G.,  2009, \aap, 502, 771

\bibitem[\protect\citeauthoryear{{Dupke} \& {White} III}{{Dupke} \&
  {White}}{2003}]{Dupke03}
{Dupke} R.,  {White} III R.~E.,  2003, \apjl, 583, L13

\bibitem[\protect\citeauthoryear{{Dupke}, {White} III \& {Bregman}}{{Dupke}
  et~al.}{2007}]{Dupke07}
{Dupke} R.,  {White} III R.~E.,    {Bregman} J.~N.,  2007, \apj, 671, 181

\bibitem[\protect\citeauthoryear{{Fabian}, {Sanders}, {Taylor}, {Allen},
  {Crawford}, {Johnstone} \& {Iwasawa}}{{Fabian} et~al.}{2006}]{fabian2006}
{Fabian} A.~C.,  {Sanders} J.~S.,  {Taylor} G.~B.,  {Allen} S.~W.,  {Crawford}
  C.~S.,  {Johnstone} R.~M.,    {Iwasawa} K.,  2006, \mnras, 366, 417

\bibitem[\protect\citeauthoryear{{Feigelson}, {Wood}, {Schreier}, {Harris} \&
  {Reid}}{{Feigelson} et~al.}{1987}]{Feigelson87}
{Feigelson} E.~D.,  {Wood} P.~A.~D.,  {Schreier} E.~J.,  {Harris} D.~E.,
  {Reid} M.~J.,  1987, \apj, 312, 101

\bibitem[\protect\citeauthoryear{{Feldman}}{{Feldman}}{1992}]{Feldman92}
{Feldman} U.,  1992, \physscr, 46, 202

\bibitem[\protect\citeauthoryear{{Finoguenov}, {David} \&
  {Ponman}}{{Finoguenov} et~al.}{2000}]{finoguenov2000}
{Finoguenov} A.,  {David} L.~P.,    {Ponman} T.~J.,  2000, \apj, 544, 188

\bibitem[\protect\citeauthoryear{{Finoguenov}, {Matsushita}, {B{\"o}hringer},
  {Ikebe} \& {Arnaud}}{{Finoguenov} et~al.}{2002}]{finoguenov2002}
{Finoguenov} A.,  {Matsushita} K.,  {B{\"o}hringer} H.,  {Ikebe} Y.,
  {Arnaud} M.,  2002, \aap, 381, 21

\bibitem[\protect\citeauthoryear{{Forman}, {Jones}, {Churazov}, {Markevitch},
  {Nulsen}, {Vikhlinin}, {Begelman}, {B{\"o}hringer}, {Eilek}, {Heinz},
  {Kraft}, {Owen} \& {Pahre}}{{Forman} et~al.}{2007}]{Forman06}
{Forman} W.,  {Jones} C.,  {Churazov} E.,  {Markevitch} M.,  {Nulsen} P.,
  {Vikhlinin} A.,  {Begelman} M.,  {B{\"o}hringer} H.,  {Eilek} J.,  {Heinz}
  S.,  {Kraft} R.,  {Owen} F.,    {Pahre} M.,  2007, \apj, 665, 1057

\bibitem[\protect\citeauthoryear{{Forman}, {Nulsen}, {Heinz}, {Owen}, {Eilek},
  {Vikhlinin}, {Markevitch}, {Kraft}, {Churazov} \& {Jones}}{{Forman}
  et~al.}{2005}]{Forman05}
{Forman} W.,  {Nulsen} P.,  {Heinz} S.,  {Owen} F.,  {Eilek} J.,  {Vikhlinin}
  A.,  {Markevitch} M.,  {Kraft} R.,  {Churazov} E.,    {Jones} C.,  2005, ApJ,
  635, 894

\bibitem[\protect\citeauthoryear{{Fujita}, {Tawa}, {Hayashida}, {Takizawa},
  {Matsumoto}, {Okabe} \& {Reiprich}}{{Fujita} et~al.}{2008}]{fujita2007}
{Fujita} Y.,  {Tawa} N.,  {Hayashida} K.,  {Takizawa} M.,  {Matsumoto} H.,
  {Okabe} N.,    {Reiprich} T.~H.,  2008, \pasj, 60, 343

\bibitem[\protect\citeauthoryear{{Iwamoto}, {Brachwitz}, {Nomoto}, {Kishimoto},
  {Umeda}, {Hix} \& {Thielemann}}{{Iwamoto} et~al.}{1999}]{iwamoto1999}
{Iwamoto} K.,  {Brachwitz} F.,  {Nomoto} K.,  {Kishimoto} N.,  {Umeda} H.,
  {Hix} W.~R.,    {Thielemann} F.,  1999, \apjs, 125, 439

\bibitem[\protect\citeauthoryear{{Kaastra}, {Mewe} \&
  {Nieuwenhuijzen}}{{Kaastra} et~al.}{1996}]{kaastra1996}
{Kaastra} J.~S.,  {Mewe} R.,    {Nieuwenhuijzen} H.,  1996, in UV and X-ray
  Spectroscopy of Astrophysical and Laboratory Plasmas p.411, K. Yamashita and
  T. Watanabe. Tokyo : Universal Academy Press {Spex: a New Code for Spectral
  Analysis of X and UV Spectra}

\bibitem[\protect\citeauthoryear{{Kalberla}, {Burton}, {Hartmann}, {Arnal},
  {Bajaja}, {Morras} \& {P{\"o}ppel}}{{Kalberla} et~al.}{2005}]{kalberla2005}
{Kalberla} P.~M.~W.,  {Burton} W.~B.,  {Hartmann} D.,  {Arnal} E.~M.,  {Bajaja}
  E.,  {Morras} R.,    {P{\"o}ppel} W.~G.~L.,  2005, \aap, 440, 775

\bibitem[\protect\citeauthoryear{{Kormendy}, {Fisher}, {Cornell} \&
  {Bender}}{{Kormendy} et~al.}{2009}]{Kormendy09}
{Kormendy} J.,  {Fisher} D.~B.,  {Cornell} M.~E.,    {Bender} R.,  2009, \apjs,
  182, 216

\bibitem[\protect\citeauthoryear{{Kronawitter}, {Saglia}, {Gerhard} \&
  {Bender}}{{Kronawitter} et~al.}{2000}]{Kronawitter00}
{Kronawitter} A.,  {Saglia} R.~P.,  {Gerhard} O.,    {Bender} R.,  2000, \aaps,
  144, 53

\bibitem[\protect\citeauthoryear{{Kuntz} \& {Snowden}}{{Kuntz} \&
  {Snowden}}{2000}]{kuntz2000}
{Kuntz} K.~D.,  {Snowden} S.~L.,  2000, \apj, 543, 195

\bibitem[\protect\citeauthoryear{{Leccardi} \& {Molendi}}{{Leccardi} \&
  {Molendi}}{2008}]{Leccardi08}
{Leccardi} A.,  {Molendi} S.,  2008, \aap, 487, 461

\bibitem[\protect\citeauthoryear{{Lieu}, {Mittaz}, {Bowyer}, {Lockman}, {Hwang}
  \& {Schmitt}}{{Lieu} et~al.}{1996}]{Lieu96}
{Lieu} R.,  {Mittaz} J.~P.~D.,  {Bowyer} S.,  {Lockman} F.~J.,  {Hwang} C.-Y.,
    {Schmitt} J.~H.~M.~M.,  1996, \apjl, 458, L5

\bibitem[\protect\citeauthoryear{{Mannucci}, {Maoz}, {Sharon}, {Botticella},
  {Della Valle}, {Gal-Yam} \& {Panagia}}{{Mannucci} et~al.}{2008}]{Mannucci08}
{Mannucci} F.,  {Maoz} D.,  {Sharon} K.,  {Botticella} M.~T.,  {Della Valle}
  M.,  {Gal-Yam} A.,    {Panagia} N.,  2008, \mnras, 383, 1121

\bibitem[\protect\citeauthoryear{{Markevitch}, {Gonzalez}, {David},
  {Vikhlinin}, {Murray}, {Forman}, {Jones} \& {Tucker}}{{Markevitch}
  et~al.}{2002}]{Markevitch02}
{Markevitch} M.,  {Gonzalez} A.~H.,  {David} L.,  {Vikhlinin} A.,  {Murray} S.,
   {Forman} W.,  {Jones} C.,    {Tucker} W.,  2002, \apjl, 567, L27

\bibitem[\protect\citeauthoryear{{Markevitch} \& {Vikhlinin}}{{Markevitch} \&
  {Vikhlinin}}{2007}]{markevitch2007}
{Markevitch} M.,  {Vikhlinin} A.,  2007, \physrep, 443, 1

\bibitem[\protect\citeauthoryear{{Markevitch}, {Vikhlinin} \&
  {Mazzotta}}{{Markevitch} et~al.}{2001}]{Markevitch01}
{Markevitch} M.,  {Vikhlinin} A.,    {Mazzotta} P.,  2001, \apjl, 562, L153

\bibitem[\protect\citeauthoryear{{Matsushita}, {Belsole}, {Finoguenov} \&
  {B{\"o}hringer}}{{Matsushita} et~al.}{2002}]{Matsushita02}
{Matsushita} K.,  {Belsole} E.,  {Finoguenov} A.,    {B{\"o}hringer} H.,  2002,
  \aap, 386, 77

\bibitem[\protect\citeauthoryear{{Matsushita}, {Finoguenov} \&
  {B{\"o}hringer}}{{Matsushita} et~al.}{2003}]{Matsushita03}
{Matsushita} K.,  {Finoguenov} A.,    {B{\"o}hringer} H.,  2003, \aap, 401, 443

\bibitem[\protect\citeauthoryear{{McNamara} \& {Nulsen}}{{McNamara} \&
  {Nulsen}}{2007}]{McNamara07}
{McNamara} B.~R.,  {Nulsen} P.~E.~J.,  2007, \araa, 45, 117

\bibitem[\protect\citeauthoryear{{Mihos}, {Harding}, {Feldmeier} \&
  {Morrison}}{{Mihos} et~al.}{2005}]{Mihos05}
{Mihos} J.~C.,  {Harding} P.,  {Feldmeier} J.,    {Morrison} H.,  2005, \apjl,
  631, L41

\bibitem[\protect\citeauthoryear{{Molendi}}{{Molendi}}{2002}]{Molendi02}
{Molendi} S.,  2002, \apj, 580, 815

\bibitem[\protect\citeauthoryear{{Nomoto}, {Tominaga}, {Umeda}, {Kobayashi} \&
  {Maeda}}{{Nomoto} et~al.}{2006}]{nomoto2006}
{Nomoto} K.,  {Tominaga} N.,  {Umeda} H.,  {Kobayashi} C.,    {Maeda} K.,
  2006, Nuclear Physics A, 777, 424

\bibitem[\protect\citeauthoryear{{Owers}, {Nulsen}, {Couch} \&
  {Markevitch}}{{Owers} et~al.}{2009}]{Owers09}
{Owers} M.~S.,  {Nulsen} P.~E.~J.,  {Couch} W.~J.,    {Markevitch} M.,  2009,
  ArXiv e-prints

\bibitem[\protect\citeauthoryear{{Peletier}, {Davies}, {Illingworth}, {Davis}
  \& {Cawson}}{{Peletier} et~al.}{1990}]{Peletier90}
{Peletier} R.~F.,  {Davies} R.~L.,  {Illingworth} G.~D.,  {Davis} L.~E.,
  {Cawson} M.,  1990, \aj, 100, 1091

\bibitem[\protect\citeauthoryear{{Peterson}, {Kahn}, {Paerels}, {Kaastra},
  {Tamura}, {Bleeker}, {Ferrigno} \& {Jernigan}}{{Peterson}
  et~al.}{2003}]{Peterson03}
{Peterson} J.~R.,  {Kahn} S.~M.,  {Paerels} F.~B.~S.,  {Kaastra} J.~S.,
  {Tamura} T.,  {Bleeker} J.~A.~M.,  {Ferrigno} C.,    {Jernigan} J.~G.,  2003,
  \apj, 590, 207

\bibitem[\protect\citeauthoryear{{Peterson}, {Paerels}, {Kaastra}, {Arnaud},
  {Reiprich}, {Fabian}, {Mushotzky}, {Jernigan} \& {Sakelliou}}{{Peterson}
  et~al.}{2001}]{Peterson01}
{Peterson} J.~R.,  {Paerels} F.~B.~S.,  {Kaastra} J.~S.,  {Arnaud} M.,
  {Reiprich} T.~H.,  {Fabian} A.~C.,  {Mushotzky} R.~F.,  {Jernigan} J.~G.,
  {Sakelliou} I.,  2001, \aap, 365, L104

\bibitem[\protect\citeauthoryear{{Rasmussen} \& {Ponman}}{{Rasmussen} \&
  {Ponman}}{2009}]{Rasmussen09}
{Rasmussen} J.,  {Ponman} T.~J.,  2009, \mnras, 399, 239

\bibitem[\protect\citeauthoryear{{Read} \& {Ponman}}{{Read} \&
  {Ponman}}{2003}]{ReadPonman}
{Read} A.~M.,  {Ponman} T.~J.,  2003, \aap, 409, 395

\bibitem[\protect\citeauthoryear{{Rebusco}, {Churazov}, {B{\"o}hringer} \&
  {Forman}}{{Rebusco} et~al.}{2006}]{Rebusco06}
{Rebusco} P.,  {Churazov} E.,  {B{\"o}hringer} H.,    {Forman} W.,  2006,
  \mnras, 372, 1840

\bibitem[\protect\citeauthoryear{{Renzini}, {Ciotti}, {D'Ercole} \&
  {Pellegrini}}{{Renzini} et~al.}{1993}]{Renzini93}
{Renzini} A.,  {Ciotti} L.,  {D'Ercole} A.,    {Pellegrini} S.,  1993, \apj,
  419, 52

\bibitem[\protect\citeauthoryear{{Sato}, {Matsushita} \& {Gastaldello}}{{Sato}
  et~al.}{2009}]{Sato09}
{Sato} K.,  {Matsushita} K.,    {Gastaldello} F.,  2009, \pasj, 61, 365

\bibitem[\protect\citeauthoryear{{Sato}, {Matsushita}, {Ishisaki}, {Sasaki},
  {Ohashi}, {Yamasaki} \& {Ishida}}{{Sato} et~al.}{2008}]{sato2007b}
{Sato} K.,  {Matsushita} K.,  {Ishisaki} Y.,  {Sasaki} S.,  {Ohashi} T.,
  {Yamasaki} N.~Y.,    {Ishida} M.,  2008, \pasj, 60, 333

\bibitem[\protect\citeauthoryear{{Sharon}, {Gal-Yam}, {Maoz}, {Filippenko} \&
  {Guhathakurta}}{{Sharon} et~al.}{2007}]{Sharon07}
{Sharon} K.,  {Gal-Yam} A.,  {Maoz} D.,  {Filippenko} A.~V.,    {Guhathakurta}
  P.,  2007, \apj, 660, 1165

\bibitem[\protect\citeauthoryear{{Simionescu}, {B\"ohringer}, Br\"{u}ggen \&
  {Finoguenov}}{{Simionescu} et~al.}{2007}]{simionescu2007a}
{Simionescu} A.,  {B\"ohringer} H.,  Br\"{u}ggen M.,    {Finoguenov} A.,  2007,
  A\&A, 465, 749

\bibitem[\protect\citeauthoryear{{Simionescu}, {Werner}, {B{\"o}hringer},
  {Kaastra}, {Finoguenov}, {Br{\"u}ggen} \& {Nulsen}}{{Simionescu}
  et~al.}{2009}]{simionescu2008b}
{Simionescu} A.,  {Werner} N.,  {B{\"o}hringer} H.,  {Kaastra} J.~S.,
  {Finoguenov} A.,  {Br{\"u}ggen} M.,    {Nulsen} P.~E.~J.,  2009, \aap, 493,
  409

\bibitem[\protect\citeauthoryear{{Simionescu}, {Werner}, {Finoguenov},
  {B{\"o}hringer} \& {Br{\"u}ggen}}{{Simionescu}
  et~al.}{2008}]{simionescu2008a}
{Simionescu} A.,  {Werner} N.,  {Finoguenov} A.,  {B{\"o}hringer} H.,
  {Br{\"u}ggen} M.,  2008, \aap, 482, 97

\bibitem[\protect\citeauthoryear{{Smith}, {Brickhouse}, {Liedahl} \&
  {Raymond}}{{Smith} et~al.}{2001}]{smith2001}
{Smith} R.~K.,  {Brickhouse} N.~S.,  {Liedahl} D.~A.,    {Raymond} J.~C.,
  2001, \apjl, 556, L91

\bibitem[\protect\citeauthoryear{{Snowden}, {Mushotzky}, {Kuntz} \&
  {Davis}}{{Snowden} et~al.}{2008}]{snowden2007}
{Snowden} S.~L.,  {Mushotzky} R.~F.,  {Kuntz} K.~D.,    {Davis} D.~S.,  2008,
  \aap, 478, 615

\bibitem[\protect\citeauthoryear{{Tamura}, {Kaastra}, {den Herder}, {Bleeker}
  \& {Peterson}}{{Tamura} et~al.}{2004}]{tamura2004}
{Tamura} T.,  {Kaastra} J.~S.,  {den Herder} J.~W.~A.,  {Bleeker} J.~A.~M.,
  {Peterson} J.~R.,  2004, \aap, 420, 135

\bibitem[\protect\citeauthoryear{{Tittley} \& {Henriksen}}{{Tittley} \&
  {Henriksen}}{2005}]{Tittley05}
{Tittley} E.~R.,  {Henriksen} M.,  2005, \apj, 618, 227

\bibitem[\protect\citeauthoryear{{Voit} \& {Donahue}}{{Voit} \&
  {Donahue}}{2005}]{Voit05}
{Voit} G.~M.,  {Donahue} M.,  2005, \apj, 634, 955

\bibitem[\protect\citeauthoryear{{Werner}, {Durret}, {Ohashi}, {Schindler} \&
  {Wiersma}}{{Werner} et~al.}{2008}]{Werner08rev}
{Werner} N.,  {Durret} F.,  {Ohashi} T.,  {Schindler} S.,    {Wiersma}
  R.~P.~C.,  2008, Space Science Reviews, 134, 337

\bibitem[\protect\citeauthoryear{{Young}, {Wilson} \& {Mundell}}{{Young}
  et~al.}{2002}]{Young02}
{Young} A.~J.,  {Wilson} A.~S.,    {Mundell} C.~G.,  2002, \apj, 579, 560

\end{thebibliography}

\label{lastpage}

\end{document}